\documentclass[aps,pra,twocolumn,superscriptaddress,nofootinbib,longbibliography,floatfix]{revtex4-2}
\usepackage{amsmath,amssymb,bm,graphicx}
\usepackage{placeins}
\usepackage[dvipsnames]{xcolor}
\usepackage{hyperref}
\usepackage{orcidlink}
\hypersetup{colorlinks=true,linkcolor=BrickRed,citecolor=RoyalBlue,urlcolor=RoyalBlue,
  pdftitle={Nonstabilizerness of the entangling layer in quantum circuit Born machines},
  pdfauthor={Marcin Plodzien}}
\graphicspath{{./}}

\newcommand{\ket}[1]{|#1\rangle}
\newcommand{\bra}[1]{\langle#1|}
\newcommand{\braket}[1]{\langle #1\rangle}
\newcommand{\KL}{\mathrm{KLD}}
\newcommand{\NLL}{\mathrm{NLL}}
\newcommand{\Us}{\hat{U}_S}
\newcommand{\NQ}{N}
\newcommand{\Mtwo}{M_2}

\begin{document}

\title{Nonstabilizerness of the entangling layer in quantum circuit Born machines}

\author{Marcin P\l{}odzie\'n\,\orcidlink{0000-0002-0835-1644}}
\affiliation{Qilimanjaro Quantum Tech, Carrer de Vene\c{c}uela 74, 08019 Barcelona, Spain}
\date{\today}

\begin{abstract}
The entangling layer of a quantum circuit Born machine can supply both entanglement and nonstabilizerness. In this work we show that the entangling layer can be a Clifford circuit, which supplies no nonstabilizerness. We use an architecture in which a fixed, non-trainable entangler is reused in every layer and only single-qubit rotations are optimized. There, a Clifford entangler reaches the same Kullback--Leibler divergence as a near-Haar one at sufficient circuit depth, and entanglers spanning zero to Haar-typical nonstabilizerness reach the same loss, up to a small offset for a free-fermion matchgate entangler. A doping control that raises the entangler's nonstabilizerness while leaving the entangling power of every gate unchanged leaves the loss flat. With a Clifford entangler the nonstabilizerness of the trained state comes entirely from the rotations, which take the trained circuit outside the stabilizer formalism. We also show that, among all pure states whose Born distribution is a given target, the one whose amplitudes are the positive square roots of the probabilities has the smallest stabilizer R\'enyi entropy, so this minimum can be computed without training.
\end{abstract}

\maketitle

\section{Introduction}\label{sec:intro}
A quantum circuit Born machine encodes a probability distribution in a parametrized quantum state measured in
the computational basis, with the Born rule assigning a normalized probability to each of the exponentially
many outcomes~\cite{liu2018differentiable,cheng2018information,coyle2020born,gao2018quantum,gao2022enhancing,kolarovszki2026gaussian}. The circuit interleaves trainable single-qubit rotations with entangling
operations~\cite{biamonte2017quantum,cerezo2021variational,benedetti2019generative,rudolph2024trainability,gili2024generalization},
and the entangling layer can supply both entanglement and
nonstabilizerness~\cite{fux2023entanglementmagic}. Clifford circuits generate
extensive entanglement but no nonstabilizerness, whereas non-Clifford single-qubit rotations generate
nonstabilizerness without coupling qubits. Nonstabilizerness, quantified here by the stabilizer R\'enyi
entropy (SRE)~\cite{leone2022,haug2023stabilizer}, separates Clifford dynamics from universal quantum
computation~\cite{aaronson2004,bravyi2005UniversalQuantumComputation} and is the resource typically supplied by
magic-state distillation in fault-tolerant architectures~\cite{veitch2014,howard2017}. It has become a standard diagnostic
of many-body
states~\cite{liu2022many,white2021,oliviero2022ising,turkeshi2023pauli,tarabunga2023manybody,niroula2023phase,Turkeshi2025}
and a probe of expressivity in variational and neural quantum
states~\cite{spriggs2024quantumresources,sinibaldi2025nonstabilizerness,ivaki2026optimal,lipardi2026magic}.
Its consumption across the gates of trained generative circuits has recently been
quantified for the IQP family~\cite{kruger2026generative}, but whether the entangling layer of a Born machine has to supply
nonstabilizerness at all, separately from entangling power, has not been tested, although fixed CNOT
entanglers, which are Clifford, are a common choice~\cite{liu2018differentiable}.

We address this question in the quantum circuit Born machine of Ref.~\cite{plodzien2026qsbm}, where a fixed
entangler is reused in every layer and only the single-qubit rotations are optimized, so that the entangler
can be exchanged with everything else held fixed. In that work, entanglers of very different microscopic
makeup all converged to the same loss once a single application reached the half-chain entanglement
of a random state, the Page value~\cite{page1993average}. All were near-Haar, and a Haar-random unitary
saturates Page entanglement, near-maximal nonstabilizerness and high moment order at once, leaving open which of these
properties drives the convergence. We find that, for the system sizes and targets studied, adding
nonstabilizerness to the entangler does not lower the generative loss.

We vary the nonstabilizerness of the entangler across random brickwork families that entangle comparably, namely
Clifford circuits, which generate none, Clifford circuits doped with a controlled number of non-Clifford
gates, and local-Haar circuits, which carry the Haar-typical amount. The
Clifford family and its doped extension are exact controls, since the SRE is invariant under Clifford
operations and doping leaves the entangling power of every gate untouched and preserves the agreement with
Haar through third order.

We first show that, among all pure states encoding a target distribution, the one whose amplitudes are the
positive square roots of the probabilities attains the least stabilizer R\'enyi entropy, which makes this minimum
exactly computable. On the Gaussian-mixture target the random entangler families reach the same loss at sufficient depth. A deterministic kicked-Ising Clifford entangler reaches a low loss on two-dimensional targets with thin,
curved or diagonal support, reproducing some of them while leaving the spiral arms unresolved. The entanglers
differ in how much entanglement a single application generates, from close to the Page value for the
Clifford, doped and local-Haar families, through about half of it for the matchgate, down to a constant for
the kicked-Ising one.

The paper is organized as follows. Section~\ref{sec:model} defines the model, the entangler families, the
nonstabilizerness diagnostics and the target distributions. Section~\ref{sec:results} gives the minimal
nonstabilizerness a distribution forces on its pure-state encodings, then the entangler comparison on the
Gaussian-mixture, spin-glass and prescribed-nonstabilizerness targets, and a demonstration on
two-dimensional targets with a single deterministic Clifford entangler. Appendix~\ref{app:proof} proves the minimum and App.~\ref{app:num} collects the numerical methods.

\section{Preliminaries}\label{sec:model}
\subsection{Model and entangler families}
\begin{figure}[t]
\centering
\includegraphics[width=\columnwidth]{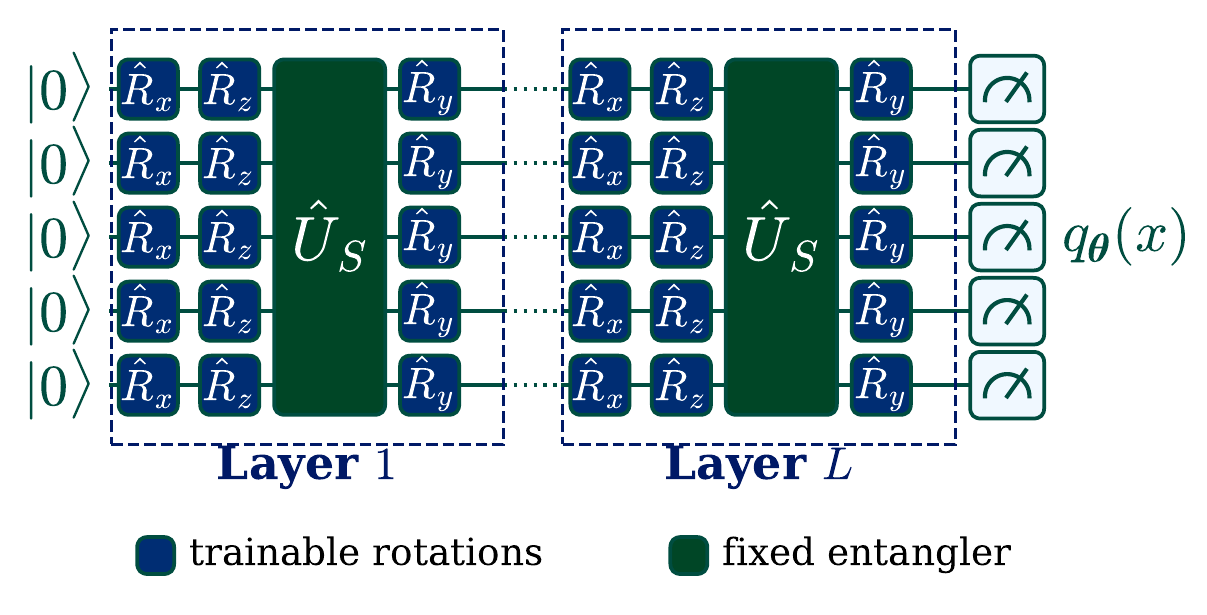}
\caption{\label{fig:circuit} The circuit of Eq.~\eqref{eq:ansatz}. Each of the $L$ layers applies trainable
rotations $\hat R_x$ and $\hat R_z$ to every qubit, the fixed entangler $\Us$ of Eq.~\eqref{eq:brickwork}, and a
trainable $\hat R_y$. The same $\Us$ enters every layer, and the register is measured in the computational
basis to give $q_{\bm\theta}(x)$.}
\end{figure}
We prepare $\NQ$ qubits in $\ket{0}^{\otimes\NQ}$ and apply $L$ identical layers (Fig.~\ref{fig:circuit}). Layer $\ell$
applies trainable single-qubit pre-rotations $\hat R_z(\theta^{(\ell)}_{i,z})\hat R_x(\theta^{(\ell)}_{i,x})$
to every qubit $i$, then the fixed entangler $\Us$, then a trainable post-rotation
$\hat R_y(\theta^{(\ell)}_{i,y})$,
\begin{equation}\label{eq:ansatz}
\hat U(\bm\theta)=\prod_{\ell=1}^{L}\Big[\textstyle\bigotimes_i\hat R_y(\theta^{(\ell)}_{i,y})\Big]\,\Us\,
\Big[\textstyle\bigotimes_i\hat R_z(\theta^{(\ell)}_{i,z})\hat R_x(\theta^{(\ell)}_{i,x})\Big],
\end{equation}
with $\hat R_\alpha(\theta)=e^{-i\theta\hat\sigma_\alpha/2}$ and $\alpha\in\{x,y,z\}$, the product ordered so that
$\ell=1$ acts first. Only the $3\NQ L$ rotation angles are trained. The output distribution is
$q_{\bm\theta}(x)=|\braket{x|\hat U(\bm\theta)|0^{\NQ}}|^2$. Throughout we stay in the under-parametrized regime
$3\NQ L<2^\NQ$, where the choice of entangler can still affect the converged loss.

The entangler is a brickwork of two-qubit gates on an open chain, of depth $K$, the number of two-qubit
sub-layers within one $\Us$, which is distinct from the layer count $L$,
\begin{equation}\label{eq:brickwork}
\Us=\prod_{k=0}^{K-1}\ \prod_{i\in B_k}\hat u^{(k)}_{i,i+1},
\end{equation}
ordered so that $k=0$ acts first, with $B_k$ the set of bonds whose left qubit index is even ($0,2,4,\dots$) for even $k$ and odd
($1,3,5,\dots$) for odd $k$, each bond acting on qubits $i$ and $i+1$. The gates $\hat u^{(k)}_{i,i+1}$ are drawn once and
held fixed, the same $\Us$ entering every layer of Eq.~\eqref{eq:ansatz}. We use five families, which
together span a range of nonstabilizerness and of moment order:
\begin{itemize}
\item \emph{Clifford}. Each gate is drawn uniformly from the two-qubit Clifford group. The entangler
generates no nonstabilizerness at any depth.
\item \emph{Doped-Clifford}. The Clifford brickwork with $n_t$ of its gates right-multiplied by $T\otimes I$,
where $T=\mathrm{diag}(1,e^{i\pi/4})$. Doping raises the nonstabilizerness in steps, reaching a probe-state SRE (Sec.~\ref{sec:diagnostics}) of $1.54$ at $n_t=4$, and since
$T\otimes I$ is a local unitary it leaves the entangling power of every gate unchanged.
\item \emph{Local-Haar}. Each gate is an independent Haar-random element of $U(4)$~\cite{mezzadri2007}. The
Haar measure applies to each two-qubit gate, and the brickwork of depth $K$ built from them only approximates
a Haar-random unitary on all $\NQ$ qubits.
\item \emph{Matchgate}. Each gate is a random nearest-neighbor matchgate, the free-fermion (Gaussian)
two-qubit gate that acts as independent $U(2)$ elements of equal determinant on the even- and odd-parity
subspaces. Matchgate circuits are classically
simulable~\cite{valiant2002,terhal2002,jozsa2008}, yet they carry
nonstabilizerness~\cite{collura2025quantummagicfermionicgaussian}.
\item \emph{Kicked-Ising}. A single deterministic gate tiled over every bond, the transverse-field
kicked-Ising bond gate at its Clifford point,
\begin{equation}\label{eq:kickedising}
\hat U_{\mathrm{bond}}=\hat R_x(\tfrac{\pi}{2})^{\otimes2}\,
   e^{-i\frac{\pi}{4}\hat\sigma_z\otimes\hat\sigma_z},
\end{equation}
whose two factors are both Clifford. It maps stabilizer states to stabilizer states and its probe-state SRE
vanishes. At $K=2\NQ$ it generates a half-chain entanglement of $2\ln2$ at every size checked.
\end{itemize}
The four random families are compared on the one-dimensional, spin-glass and prescribed-nonstabilizerness
targets, where the doped family sweeps the nonstabilizerness in steps at fixed entangling power and
carries the control. The kicked-Ising family enters the two-dimensional targets.

The Clifford, doped and local-Haar gate ensembles agree in all moments up to third order and first differ
at fourth order, the order at which the SRE enters. The two-qubit Clifford group is a unitary
$3$-design~\cite{webb2016,Gross2021}, and right-multiplying a gate by the fixed $T\otimes I$ leaves its first
three moments unchanged. We call the highest $t$ for which a gate ensemble is a unitary $t$-design its moment
order. The matchgate ensemble forms no unitary design at any order.

Unless stated otherwise the entangler depth is $K=2\NQ$, where a single application of the Clifford, doped or
local-Haar entangler generates close to the Page value of half-chain entanglement at every size checked, up
to $\NQ=20$, so that these three families differ in little besides nonstabilizerness. The matchgate family differs from the others in all three respects at once, in nonstabilizerness, in
moment order, and in the rate at which it generates entanglement.

\subsection{Loss function and diagnostics}\label{sec:diagnostics}
We train the model to minimize the Kullback--Leibler divergence to the target,
\begin{equation}\label{eq:kld}
\KL(p\Vert q_{\bm\theta})=\sum_x p(x)\ln\frac{p(x)}{q_{\bm\theta}(x)}.
\end{equation}
Since $\KL=\NLL-S(p)$, with $\NLL$ the negative log-likelihood and $S(p)$ the fixed Shannon entropy of the
target in nats, this is equivalent to maximizing the likelihood. We refer to $\KL$ as the loss throughout.
It is computed exactly from the statevector probabilities
$q_{\bm\theta}(x)$, with no sampling contribution.
Each family is trained $20$ times at the same $(\NQ,K,L)$. The runs start from different random angles,
and each run of a random family draws a new entangler. All families share the same $20$ starting angles and
epoch budget, so run $i$ of one family is paired with run $i$ of every other. For the spin glass we use $10$ disorder realizations with $10$ runs each. Statistics and optimizer settings are given in
App.~\ref{app:num}.

We quantify the nonstabilizerness of a \emph{state} by the stabilizer R\'enyi
entropy~\cite{leone2022,haug2023stabilizer,tarabunga2023manybody,Sierant_2026},
\begin{equation}\label{eq:sre}
\Mtwo(\psi)=-\log_2\!\Big[\tfrac{1}{2^\NQ}\!\sum_{P\in\mathcal P_\NQ}\!\braket{\psi|P|\psi}^4\Big],
\end{equation}
where $\mathcal P_\NQ$ is the set of $4^{\NQ}$ Hermitian Pauli strings on $\NQ$ qubits. It is non-negative, vanishes if and only if $\psi$ is a
stabilizer state, and approaches $\NQ-2$ for a typical Haar-random state~\cite{leone2022}. We evaluate it
exactly in $O(\NQ\,4^\NQ)$ operations, one fast Walsh--Hadamard transform of cost $O(\NQ\,2^{\NQ})$ in each
of the $2^{\NQ}$ Pauli sectors~\cite{Sierant_2026}. We characterize an entangler by the SRE of its output on the fixed reference input
$\ket{+}^{\otimes\NQ}$, with $\ket{+}=(\ket{0}+\ket{1})/\sqrt2$, calling
$\Mtwo(\Us\ket{+}^{\otimes\NQ})$ the probe-state SRE. It is a single number attached to $\Us$, and its
value depends on the choice of input, since a non-Clifford unitary can map one stabilizer state to another
and so return zero. Within the doped family the control variable is $n_t$, the probe-state SRE grows with
$n_t$, and we report both (Sec.~\ref{sec:results}). Across families we use the probe-state SRE to order
them.

\subsection{Target probability distributions}\label{sec:targets}
As the primary one-dimensional target we use a five-peak Gaussian mixture on $2^\NQ$ bins
[shaded in Fig.~\ref{fig:combined}(a)], $p(x)\propto\sum_{j=1}^{5}w_j\exp[-(x-\mu_j)^2/2\sigma^2]$, which is smooth and of low intrinsic
nonstabilizerness. The means are uniformly spaced at $\mu_j=(j-\tfrac12)2^\NQ/5$, the common width is
$\sigma=2^\NQ/20$, and the weights are drawn once from the uniform distribution on $[0.2,2.0]$. The bin index maps to the
computational basis in standard binary order, with qubit $0$ carrying the most significant bit.

The second target is the Gibbs distribution of a frustrated random spin glass. We use the Sherrington--Kirkpatrick (SK) model~\cite{sherrington1975solvable}
of $\NQ$ Ising spins $s_i=\pm1$,
\begin{equation}\label{eq:sk}
H(s)=\sum_{i<j}J_{ij}\,s_i s_j,\qquad J_{ij}=\frac{g_{ij}}{\sqrt{\NQ}},\quad g_{ij}\sim\mathcal N(0,1),
\end{equation}
the standard SK normalization. With each spin configuration identified with a computational-basis
state, the target is the Gibbs distribution at inverse temperature $\beta$,
\begin{equation}\label{eq:gibbs}
p_\beta(x)=e^{-\beta H(x)}/Z_\beta,\qquad Z_\beta=\textstyle\sum_x e^{-\beta H(x)}.
\end{equation}
The couplings are drawn once per disorder realization, reproducibly from the disorder index, and the same energies are reweighted at every $\beta$, which keeps the whole temperature scan on the same $10$ disorder realizations.

In two dimensions, two coordinates are encoded in two $\NQ/2$-qubit registers
($\NQ=10$, a $32\times32$ grid), and we take six targets: a mixture of four
Gaussian lobes, two concentric rings, two interleaving moons, three diagonal stripes, a diagonal cross, and
two interleaving spirals, shown in Fig.~\ref{fig:twoddist}. All but the four-lobe mixture
have thin, curved or diagonal support that an output factorizing over the two registers cannot capture.
Qubits $0,\dots,\NQ/2-1$ carry the register integer $m_u$ and the remaining qubits carry $m_v$, and the basis string
$x$ has flat index $m_u\,2^{\NQ/2}+m_v$. The two registers map to coordinates $u$ and $v$, each spanning $[-3,3]$
in $2^{\NQ/2}$ equally spaced points. The four-lobe mixture sums isotropic Gaussians of width $0.5$ centered
at $(u,v)=(\pm1.5,\pm1.5)$. The other five are Gaussian ridges of width $\delta$ about a curve,
$p\propto e^{-d^2/2\delta^2}$ with $d$ the Euclidean distance from the grid point $(u,v)$ to the curve, with the ridges of separate components summed for the rings, stripes and cross. The
curves are two concentric circles of radii $1.0$ and $2.3$ at $\delta=0.26$; the two crescents of the
standard two-moons geometry at radius $2$ and $\delta=0.30$, recentered on the grid; three diagonal lines
$u-v\in\{-2.5,0,2.5\}$ at $\delta=0.35$; the pair of diagonals $u-v=0$ and $u+v=0$ at $\delta=0.28$; and two
Archimedean spirals $r=a\vartheta/2\pi n_{\mathrm{turn}}$ with $a=2.7$, $n_{\mathrm{turn}}=2.5$ and
$\vartheta\in[0.35,2\pi n_{\mathrm{turn}}]$, the second obtained from the first by a rotation through $\pi$,
at $\delta=0.18$.

\section{Results}\label{sec:results}
\subsection{Intrinsic nonstabilizerness of a target distribution}\label{sec:m2min}
Every state
\begin{equation}\label{eq:bornrep}
\ket{\psi_\varphi}=\sum_x\sqrt{p_x}\,e^{i\varphi_x}\ket{x},\qquad \varphi_x\in[0,2\pi),
\end{equation}
has the Born distribution $p$. Setting $\varphi_x\equiv0$ gives the positive square-root state
\begin{equation}\label{eq:sqrtp}
\ket{\sqrt p}=\sum_x\sqrt{p_x}\,\ket{x},
\end{equation}
the Rokhsar--Kivelson wavefunction of $p$. The construction goes back to the equal-amplitude dimer
superposition of Ref.~\cite{rokhsar1988} and was extended to an arbitrary classical weight in
Refs.~\cite{henley2004,ardonne2004,castelnovo2005smf}. The nonstabilizerness of such states was studied in
Ref.~\cite{tarabunga2024rk}. Multiplying
Eq.~\eqref{eq:sqrtp} by a diagonal unitary $\sum_x e^{i\varphi_x}\ket{x}\bra{x}$ reproduces
Eq.~\eqref{eq:bornrep} and leaves every probability $p_x$ untouched, so Eq.~\eqref{eq:bornrep} is the set of all
pure states whose Born distribution is $p$. We call this set the phase orbit of $p$, and the
nonstabilizerness varies across it. The minimum
\begin{equation}\label{eq:m2min}
M_{2,\min}(p)=\min_{\{\varphi_x\}}M_2\!\left(\ket{\psi_\varphi}\right)
\end{equation}
is a property of $p$ alone, and $M_{2,\min}(p)>0$ implies that every pure-state encoding of $p$ carries
nonstabilizerness. The value of $M_{2,\min}(p)$ depends on the computational basis in which $p$ is
presented. Although the minimization is non-convex, the minimum is attained in closed form at $\varphi_x\equiv0$, that is, by the positive
square-root state,
\begin{equation}\label{eq:m2sqrt}
M_{2,\min}(p)=M_2\!\left(\ket{\sqrt p}\right).
\end{equation}
The phases enter the Pauli expectations only through the complex argument of
$\overline{\psi_x}\psi_{x\oplus a}$, whose modulus $\sqrt{p_x p_{x\oplus a}}$ is fixed by $p$. Bounding each
term of the Pauli fourth moment by its modulus gives a bound fixed by $p$ alone, and $\varphi_x\equiv0$ makes
every term real and non-negative at once, saturating that bound in all $2^\NQ$ Pauli sectors simultaneously.
(See App.~\ref{app:proof}.) Every
diagonal Clifford phase pattern attains the same value, and the same argument minimizes the order-$\alpha$ stabilizer
R\'enyi entropy $M_\alpha$~\cite{leone2022} at every
integer $\alpha\ge2$. $M_{2,\min}(p)$ is exactly computable at the cost of a single SRE
evaluation, and is the reference used for each target below.

The quantity is the phase-orbit analogue of the nonlocal nonstabilizerness of
Ref.~\cite{cao2025gravitational}, the part that no local unitary can remove, and
Ref.~\cite{chu2026phase} studies the counterpart in which only the $\NQ$ single-qubit phases vary. The
nonstabilizerness of $\ket{\sqrt p}$ itself has been computed for classical models admitting a stochastic
matrix form decomposition~\cite{tarabunga2024rk}, and Eq.~\eqref{eq:m2sqrt} identifies that value as the
minimum over the orbit. The states $\sum_x s_x\sqrt{p_x}\ket{x}$ with $s_x=\pm1$ form the real sign subfamily of
Eq.~\eqref{eq:bornrep}, whose SRE becomes extensive for random signs when $p$ is broad~\cite{piemontese2023}, while the
sign-free choice $s_x\equiv+1$ is the point of the same orbit at which the SRE is smallest.

\subsection{One-dimensional Gaussian mixture}\label{sec:cliffordcontrol}
The five-peak Gaussian mixture at $\NQ=10$ has intrinsic nonstabilizerness, or floor,
$M_{2,\min}=0.56$. The nonstabilizerness the trained circuit develops is that of a random-phase encoding of its own
output, as Sec.~\ref{sec:spinglass} shows.

\begin{figure}[t]
\centering
\includegraphics[width=0.98\columnwidth]{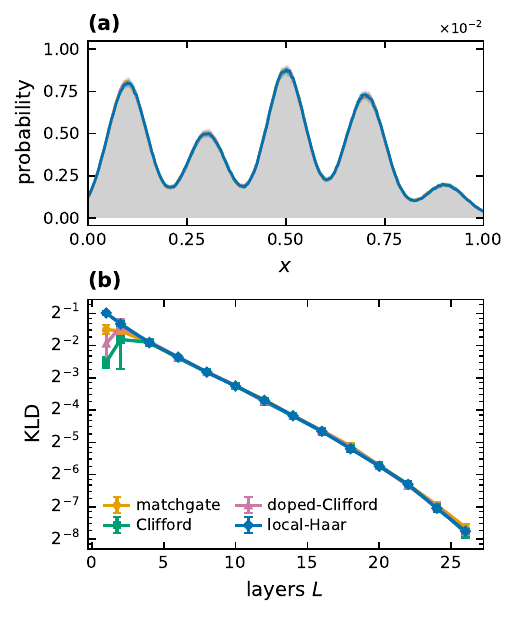}
\caption{\label{fig:combined} Panel (a) shows the five-peak Gaussian-mixture target (shaded) together with
the trained Born output at $L=26$ and $\NQ=10$ for each of the four entanglers, coarse-grained to $256$ bins (mean over $20$ runs;
band, one standard deviation). The four outputs lie on top of one another and on the target. Panel (b)
shows the converged $\KL(p\Vert q_{\bm\theta})$ against depth $L$ at $\NQ=10$, $K=2\NQ$ (mean and standard
deviation over $20$ runs; doped-Clifford at $n_t=4$), where the four curves overlap within the run-to-run spread for $L\ge4$.}
\end{figure}

With a Clifford entangler the trained state's nonstabilizerness can only come from the rotations, and that
restriction costs nothing. When the probe-state SRE of the entangler is varied from
zero (Clifford) through about $6.5$ (matchgate) to the Haar-typical $\NQ-2$ (local-Haar), the converged
$\KL$ barely changes. The four
families produce outputs indistinguishable at the resolution of Fig.~\ref{fig:combined}(a), and their
$\KL$-versus-depth curves overlap within the run-to-run spread for $L\ge4$
[Fig.~\ref{fig:combined}(b)]. The same
holds at $\NQ=8$ over $L\le10$, where the parameter count stays within $3\NQ L<2^\NQ$. At $\NQ=12$ the
families also coincide over $6\le L\le10$ and at $L=22$, $24$ and $26$, the last below the loss of the best
product distribution given below. At $L\le2$ the families separate. The paired
differences from the Clifford entangler at
$\NQ=10$, formed on the shared initial angles and target, are consistent with zero for the doped and
local-Haar families at every depth from $L=4$ on, each within about one and a half standard errors and small
against the loss itself. The
matchgate difference is positive at every depth from $L=12$ to $L=26$, reaching about four standard errors
at the largest, the same small offset it carries on the spin glass and on the targets of
Sec.~\ref{sec:m2scan}.

An entangler that generates less than the Page value in a single application reaches it over successive
layers. The
matchgate reaches only about half the Page value in one application, yet after training at $L=10$ the
half-chain entropy of the output is close to the Page value for all four families.

\begin{figure}[t]
\centering
\includegraphics[width=\columnwidth]{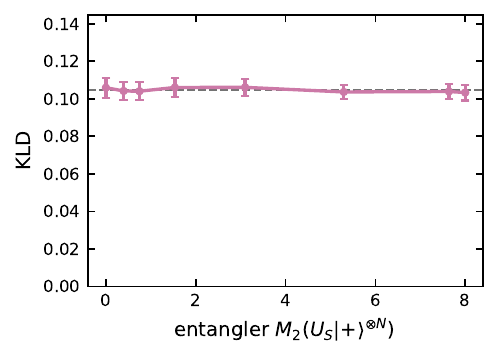}
\caption{\label{fig:doping} Converged $\KL$ against the probe-state SRE $\Mtwo(\Us\ket{+}^{\otimes\NQ})$
of a $T$-doped Clifford entangler on the five-peak Gaussian mixture, tuned by the number $n_t$ of doped
gates at fixed entangling power
($\NQ=10$, $K=2\NQ$, $L=10$; mean and standard deviation over $20$ runs; dashed, overall mean).}
\end{figure}

The $T$-doping sweep isolates the nonstabilizerness at fixed entangling power, each doping level averaged
over its own Clifford draws. Doping from $n_t=0$ to $n_t=90$, every gate of the $K=2\NQ$ brickwork, carries the
probe-state SRE from zero to the Haar-typical $\NQ-2=8$ and leaves the converged loss flat
(Fig.~\ref{fig:doping}); from end to end of the sweep the loss moves by less than the scatter of a
single point.

The product-state bound sets the scale of the nonstabilizerness at issue. Rotation layers with no entangler
between them compose to a single product state at any circuit depth $L$, and a product state carries an SRE
of at most $\NQ\log_2(3/2)\approx0.585\,\NQ$, which is $5.85$ at $\NQ=10$. Every target used here and in
Secs.~\ref{sec:spinglass} and \ref{sec:analog} has an intrinsic nonstabilizerness below that number. The bound constrains no state the trained circuit can reach, since a product state encodes none of these
targets. What a product state does reach is the closest product distribution to the target, the product of
its marginals, at $\KL=0.15$ on the Gaussian mixture. Every entangler in Fig.~\ref{fig:combined}(b) is more
than an order of magnitude below that by $L=26$.

\subsection{Spin-glass Gibbs targets}\label{sec:spinglass}
\begin{figure}[t]
\centering
\includegraphics[width=0.92\columnwidth]{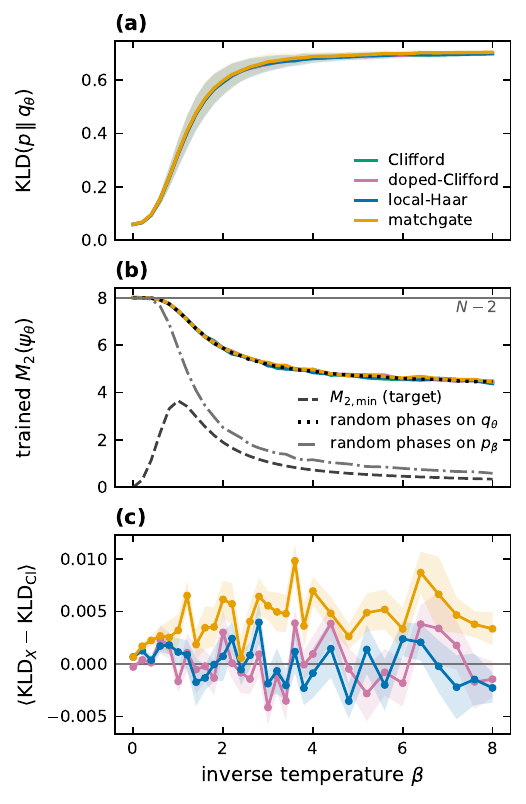}
\caption{\label{fig:gibbskld} All three panels follow the entangler families on the SK Gibbs target against
inverse temperature $\beta$ ($\NQ=10$, $K=2\NQ$, $L=12$, $10$ disorder realizations, the median over the
$10$ runs of a disorder realization taken first in panels (a) and (c); bands, one standard deviation in
panel (a)). Panel (a) shows the converged $\KL$ of each family; the curves coincide to within the line
width; panel (c) resolves the matchgate offset. Panel (b) shows the SRE of the trained state
(lowest-loss run) against the target floor $M_{2,\min}$ and two random-phase references, obtained by assigning
independent uniform phases to $\sqrt{q_{\bm\theta}}$ and to $\sqrt{p_\beta}$ for the Clifford entangler
(averaged over disorder and three phase draws), with the Haar-typical $\NQ-2$ as a thin solid line. Panel (c) shows the paired difference
$\langle \KL_X-\KL_{\mathrm{Cl}}\rangle$ over disorder (band, standard error; colors as in panel (a)), which
fluctuates about zero for the doped and local-Haar families.}
\end{figure}

Temperature tunes the intrinsic nonstabilizerness of the SK Gibbs targets. $M_{2,\min}(p_\beta)$
(Sec.~\ref{sec:m2min}) vanishes at both extremes, the uniform distribution at $\beta\to0$ and the
ground-state pair at $\beta\to\infty$ both having stabilizer square-root states, and peaks near the freezing
scale $\beta_c=1$, where it reaches $M_{2,\min}\approx3.65$ at $\NQ=10$ [dashed line in Fig.~\ref{fig:gibbskld}(b)].

The loss rises toward a plateau as $\beta$ grows [Fig.~\ref{fig:gibbskld}(a)], since with only $3\NQ L$
parameters the fixed-depth circuit does not capture the increasingly frozen distribution.

The paired difference $\langle \KL_X-\KL_{\mathrm{Cl}}\rangle$, with $X$ the doped, local-Haar or matchgate
family and $\mathrm{Cl}$ the Clifford one, cancels the large target-to-target variation
[Fig.~\ref{fig:gibbskld}(c)]. The doped and local-Haar entanglers fluctuate about zero there with no trend
in $\beta$, including across the peak of $M_{2,\min}$. Their point-to-point scatter bounds any systematic
dependence on the entangler's nonstabilizerness at the same level.

The matchgate loss is higher by a few parts in a thousand over most of the temperature range, the only
difference between the families that persists across the whole temperature range. That offset shrinks in absolute size as the circuit deepens from $L=12$ to
$L=30$ at each of the three temperatures run at both depths. The local-Haar entangler carries the most
nonstabilizerness of the four and shows no offset, so the ordering does not follow the entangler's
nonstabilizerness.

The trained-state SRE [Fig.~\ref{fig:gibbskld}(b)] exceeds the target floor at every temperature, by an
order of magnitude at the lowest one. Assigning independent
uniformly random phases to $\sqrt{q_{\bm\theta}(x)}$ reproduces the trained-state SRE at every temperature, while the same
random-phase prescription applied to $\sqrt{p_\beta(x)}$ gives a value below it once the target freezes [dotted and dash-dotted lines in
Fig.~\ref{fig:gibbskld}(b)]. The loss fixes the output probabilities but leaves the amplitude phases free, and the trained state then
carries the nonstabilizerness of a random-phase encoding of $q_{\bm\theta}$.

\begin{figure}[t]
\centering
\includegraphics[width=\columnwidth]{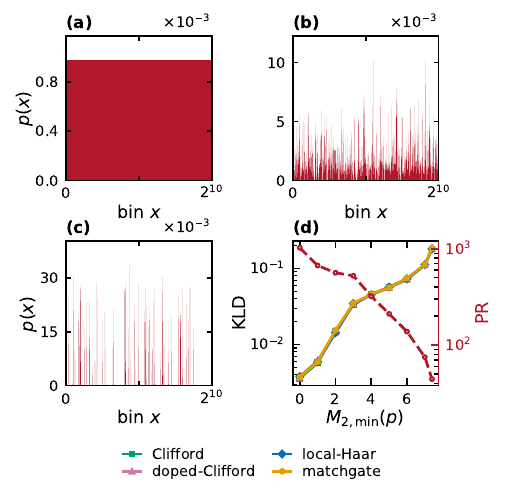}
\caption{\label{fig:m2scan} Panels (a)--(c) show the target probability $p(x)$ for three of the nine
targets, at achieved intrinsic nonstabilizerness $M_{2,\min}(p)=0$, $4.00$ and $7.40$, with
participation ratios $1024$, $321$ and $44$. Panel (d) shows the converged $\KL$ against the achieved floor
for the four entangler families on the left axis ($\NQ=10$, $K=2\NQ$, $L=26$; median over $20$ runs,
bands, the interquartile range), and the participation ratio of the same targets on the right axis (dashed).
The four families lie on top of one another, while the two axes move in opposite
directions.}
\end{figure}
\begin{figure}[t]
\centering
\includegraphics[width=0.90\columnwidth]{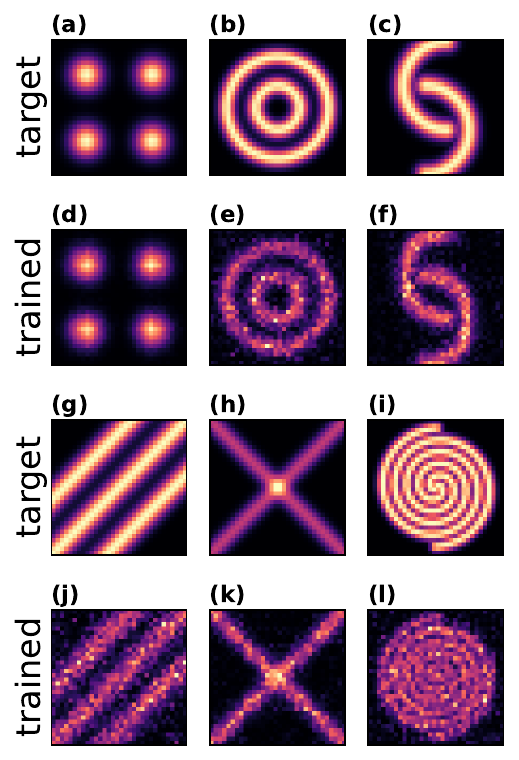}
\caption{\label{fig:twoddist} Kicked-Ising entangler (vanishing SRE) on six two-dimensional targets
($\NQ=10$, $K=2\NQ$, $L=12$, depth curriculum). Each target sits directly above its trained Born output for the
lowest-loss run, so rows one and three are targets and rows two and four the trained outputs. The
six targets are four lobes, two rings and two moons in the upper block, and three stripes, a diagonal cross
and two spirals in the lower one. Each panel is scaled to its own maximum, so brightness is
comparable within a panel but never between panels. The four-lobe mixture is reproduced; the curved and diagonal
targets are speckled along their support, and the spiral arms are not resolved.}
\end{figure}
\begin{figure}[t]
\centering
\includegraphics[width=0.94\columnwidth]{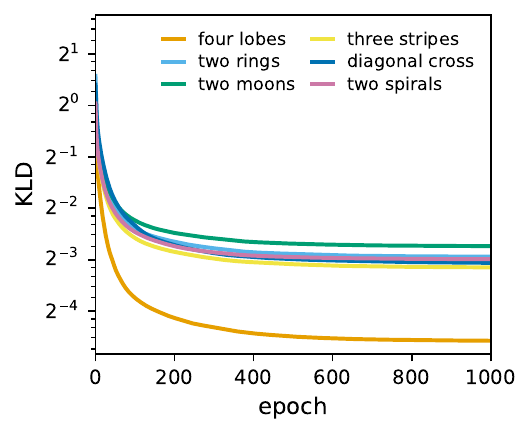}
\caption{\label{fig:twodfam} The $\KL$ through the final curriculum stage $L=12$ for the kicked-Ising
entangler, one line per two-dimensional target ($\NQ=10$, $K=2\NQ$; mean over $20$ runs).}
\end{figure}
\subsection{Targets at prescribed nonstabilizerness}\label{sec:m2scan}
Every target used so far demands less nonstabilizerness than a product state can carry
(Sec.~\ref{sec:cliffordcontrol}). To test the
comparison where that ceases to hold, we build distributions at prescribed floors by minimizing
$(M_{2,\min}(p)-m^{*})^{2}$ over $p$, with $M_{2,\min}(p)=\Mtwo(\ket{\sqrt p})$ and $m^{*}$ the requested
floor, taking nine targets from the uniform distribution up to $7.40$, the largest floor this optimization
reached at $\NQ=10$. The targets are built directly as probability vectors, with no circuit involved, and
the same $p$ is given to every entangler.

Figure~\ref{fig:m2scan}(d) gives the converged $\KL$ at $L=26$ against the achieved floor. The loss rises by a
factor of $50$ across the range. The targets also grow sparser as their floor rises
[Fig.~\ref{fig:m2scan}(a)--(c)], and their participation ratio $\mathrm{PR}=1/\sum_x p(x)^2$, on the right
axis of panel (d), falls by more than an order of magnitude, so this rise may reflect sparsity as much as
nonstabilizerness. The four families stay together throughout, their spread comparable to the
run-to-run scatter on every target. The doped and local-Haar differences from the Clifford entangler are
consistent with zero, and the matchgate carries the same small positive offset it shows on the spin glass. The three highest floors lie above the $0.585\NQ=5.85$ of Sec.~\ref{sec:cliffordcontrol}, beyond the reach of the
rotations without an entangler; the families coincide there as well.

\subsection{Two-dimensional targets}\label{sec:analog}
We turn to two-dimensional targets with thin, curved or diagonal support. Their intrinsic
nonstabilizerness runs from $1.25$ for the four-lobe mixture to $2.14$ for the two moons, well below the
product-state bound of Sec.~\ref{sec:cliffordcontrol}. Figure~\ref{fig:twoddist} uses
the kicked-Ising entangler of Eq.~\eqref{eq:kickedising}, tiled over the same brickwork of depth $K=2\NQ$.
Its probe-state SRE vanishes.

Random initial angles stall the optimization for $L\gtrsim6$ on these targets, so we train with a depth
curriculum, $L=1,2,4,\dots,12$, in which most runs at each depth start from the best angles of the previous
depth (App.~\ref{app:num}). Figure~\ref{fig:twoddist} shows the lowest-loss run of the $20$.

At $L=12$ the lowest-loss run of every target reaches $\KL<0.13$. The four-lobe mixture is reproduced and the diagonal cross and
the rings keep their shape, while the
arms of the spirals are not resolved at all, so the loss threshold is met on targets whose
structure the output reproduces and on targets where it does not. The loss flattens before the epoch
budget ends on every target (Fig.~\ref{fig:twodfam}). As in one dimension, the trained-state SRE
exceeds the target floor by a factor of three or more on every target.

\FloatBarrier
\section{Discussion and conclusions}\label{sec:discussion}
We studied whether the entangling layer of a quantum circuit Born machine must supply nonstabilizerness,
using an architecture in which a fixed, non-trainable entangler is reused in every layer while only the
single-qubit rotations are optimized, at up to $\NQ=12$ qubits. Across smooth Gaussian mixtures, frustrated spin-glass Gibbs
distributions, targets built at prescribed nonstabilizerness, and $T$-doping at fixed entangling power, the
converged loss is insensitive to the entangler's own nonstabilizerness. A Clifford entangler, whose nonstabilizerness
vanishes exactly, reaches the loss of a near-Haar one, including on targets whose intrinsic demand
exceeds what any product state can carry. A deterministic Clifford entangler also reaches a low loss on
two-dimensional targets with thin, curved or diagonal support, reproducing some of them while leaving the spiral
arms unresolved. Since the Clifford and doped gate ensembles agree with the local-Haar one only through third
order, the loss is equally insensitive to the higher moments.

The entangling layer accordingly need not supply nonstabilizerness: a Clifford brickwork serves as well
as a near-Haar one once the circuit depth $L$ is sufficient in one dimension, and a matchgate brickwork carries
a small positive offset that persists to the largest depth run and whose absolute size shrinks on the spin glass
as $L$ grows from $12$ to $30$. The trained circuit nevertheless lies
outside the stabilizer formalism, since its interleaved rotations are non-Clifford.

We also showed that the least stabilizer R\'enyi entropy a target distribution forces on any pure-state
encoding of it is attained by the positive square-root state. That replaces a non-convex minimization over $2^{\NQ}$ phases by
a single SRE evaluation, and makes the intrinsic nonstabilizerness of a distribution computable before any training.

\textit{Data availability.} The code and the data underlying every figure are openly available in
Ref.~\cite{github_repo}.

\begin{acknowledgments}
We thank Timothy Heightman for reading the manuscript and for useful comments. We acknowledge RES resources provided by Barcelona Supercomputing Center in MareNostrum 5 to NNO-2025-3-0004, and funds from MICIU/AEI/10.13039/501100011033/ FEDER, UE.
\end{acknowledgments}

\appendix

\section{Minimal nonstabilizerness of a Born distribution}\label{app:proof}
The minimum Eq.~\eqref{eq:m2min} of $\Mtwo$ over the phase orbit Eq.~\eqref{eq:bornrep} of a distribution $p$ is
attained by the positive square-root state, Eq.~\eqref{eq:m2sqrt}.

We write $a,b,x\in\{0,1\}^{\NQ}$ for bit strings of length $\NQ$, with $x$ labeling the
computational-basis state $\ket{x}$. Two strings are combined by the bitwise exclusive-or
$x\oplus a$, which flips the bits of $x$ selected by $a$, for instance $0110\oplus0011=0101$, and by the
overlap parity $a\cdot b=\sum_i a_ib_i$ taken modulo two, which is $0$ or $1$ according to whether $a$ and
$b$ share an even or an odd number of set bits.

A Pauli string is a tensor product of $\NQ$ single-qubit operators drawn from
$\{\hat I,\hat\sigma_x,\hat\sigma_y,\hat\sigma_z\}$, and the $4^{\NQ}$ of them are labeled by a pair of bit
strings through
\begin{equation}
X^a=\bigotimes_{i=1}^{\NQ}\hat\sigma_x^{\,a_i},\qquad
Z^b=\bigotimes_{i=1}^{\NQ}\hat\sigma_z^{\,b_i},
\end{equation}
so that $\hat\sigma_x$ acts on the qubits where $a_i=1$ and the identity on the rest, and likewise for
$\hat\sigma_z$ and $b$. Acting on a basis state,
\begin{equation}
X^a\ket{x}=\ket{x\oplus a},\qquad Z^b\ket{x}=(-1)^{b\cdot x}\ket{x},
\end{equation}
so $a$ selects the qubits that are flipped and $b$ those that acquire a sign. Because $\hat\sigma_y=i\hat\sigma_x\hat\sigma_z$,
every Pauli string equals $X^aZ^b$ up to a phase, and the choice $P_{a,b}=i^{a\cdot b}X^aZ^b$ is Hermitian,
so the $4^{\NQ}$ pairs $(a,b)$ enumerate $\mathcal P_\NQ$ up to sign.

The argument runs through the Walsh--Hadamard transform, the Fourier transform for functions of bit strings.
Such a function assigns one complex number to each of the $2^{\NQ}$ strings. The amplitudes $\psi_x$ of a
state are one example, and the functions $c_a$ defined below are the ones the proof transforms. The transform
sends $f$ to
\begin{equation}\label{eq:walsh}
\widehat f(b)=\sum_x(-1)^{b\cdot x}f(x),
\end{equation}
in which the sum runs over all $2^{\NQ}$ bit strings $x$ and $\widehat f$ is again a function of bit
strings. By using the orthogonality relation
\begin{equation}\label{eq:orth}
\sum_b(-1)^{b\cdot y}=2^{\NQ}\,\delta_{y,0},
\end{equation}
we turn every sum over $b$ into a constraint on the bit strings that remain. The relation holds because for
$y\neq0$ some bit $y_j$ equals one, and pairing each $b$ with $b\oplus e_j$, where $e_j$ carries a single set
bit in position $j$, reverses the sign of the summand so that the two terms cancel. A fast implementation of the transform costs $O(\NQ\,2^{\NQ})$
operations, following Ref.~\cite{Sierant_2026}, and this is what makes $\Mtwo$ tractable at the sizes
studied here.

Since $\Mtwo=-\log_2[2^{-\NQ}\sum_P\braket{P}^4]$ and $-\log_2$ is decreasing, minimizing $\Mtwo$ over the
phases is the same as maximizing the Pauli fourth moment
$\Sigma(\varphi)=\sum_{P}\braket{\psi_\varphi|P|\psi_\varphi}^4$.

For the state Eq.~\eqref{eq:bornrep} the expectation of a Pauli string is
\begin{equation}\label{eq:pexp}
\begin{split}
\braket{\psi_\varphi|P_{a,b}|\psi_\varphi}&=(-i)^{a\cdot b}\!\sum_x(-1)^{b\cdot x}\,\overline{\psi_{x}}\,\psi_{x\oplus a}\\
&=(-i)^{a\cdot b}\,\widehat{c_a}(b),
\end{split}
\end{equation}
where
\begin{equation}\label{eq:ca}
c_a(x)=\overline{\psi_x}\,\psi_{x\oplus a}=\sqrt{p_x p_{x\oplus a}}\;e^{i(\varphi_{x\oplus a}-\varphi_x)}.
\end{equation}
The $2^{\NQ}$ expectations that share the same $a$ are thus the Walsh transform of the single function
$c_a$, and the Pauli sum separates into $2^{\NQ}$ sectors labeled by $a$. Each $\braket{P_{a,b}}$ is real
because $P_{a,b}$ is Hermitian, so $\braket{P_{a,b}}^4=|\widehat{c_a}(b)|^4$ and the prefactor $(-i)^{a\cdot b}$
drops out.

Equation~\eqref{eq:ca} is what carries the proof, because the phases enter $c_a$ only through its complex
argument, while its modulus $|c_a(x)|=\sqrt{p_x p_{x\oplus a}}$ is fixed by $p$.

Writing $|\widehat{c_a}(b)|^4=\widehat{c_a}(b)^2\,\overline{\widehat{c_a}(b)}^2$, expanding both squares and
summing over $b$ with the orthogonality relation gives
\begin{equation}\label{eq:fourth}
\sum_b\big|\widehat{c_a}(b)\big|^4
 =2^{\NQ}\sum_{x_1,x_2,x_3}
 c_a(x_1)\,\overline{c_a(x_2)}\,c_a(x_3)\,\overline{c_a(x_4)},
\end{equation}
in which the three sums are unrestricted and the fourth string is fixed by
$x_4=x_1\oplus x_2\oplus x_3$, which is what Eq.~\eqref{eq:orth} leaves behind when the sum over $b$ is
carried out.
The left-hand side of Eq.~\eqref{eq:fourth} is a sum of fourth powers of moduli, hence real and
non-negative, and a real non-negative number equals its own modulus. Taking the modulus of the right-hand
side therefore changes nothing, after which the triangle inequality $|\sum_k z_k|\le\sum_k|z_k|$ applies to
the sum term by term. Each term factorizes into moduli of individual $c_a$, and by Eq.~\eqref{eq:ca} those
moduli carry no phase,
\begin{equation}\label{eq:termmod}
\big|c_a(x_1)\overline{c_a(x_2)}c_a(x_3)\overline{c_a(x_4)}\big|
=\prod_{i=1}^{4}\sqrt{p_{x_i}p_{x_i\oplus a}}.
\end{equation}
Every sector therefore obeys
\begin{equation}\label{eq:bound}
\sum_b\big|\widehat{c_a}(b)\big|^4\;\le\;2^{\NQ}\sum_{x_1,x_2,x_3}\;
 \prod_{i=1}^{4}\sqrt{p_{x_i}p_{x_i\oplus a}},
\end{equation}
in which the phases $\varphi$ have disappeared and the right-hand side is fixed by $p$ alone.

The triangle inequality is saturated when every term is non-negative. Setting $\varphi_x\equiv0$ makes
$\psi_x=\sqrt{p_x}\ge0$ and hence $c_a(x)=\sqrt{p_x p_{x\oplus a}}\ge0$ for every $a$ and every $x$, so this
single choice saturates Eq.~\eqref{eq:bound} in all $2^{\NQ}$ sectors simultaneously. It therefore maximizes
$\Sigma$ and minimizes $\Mtwo$, which is Eq.~\eqref{eq:m2sqrt}. For amplitudes that are flat on their
support this was proved in Ref.~\cite{liu2026cws} by the same argument; the derivation above extends it to
arbitrary $p$.

Many phase patterns attain the minimum, since any pattern realizable by a diagonal Clifford
leaves $\Mtwo$ invariant; the minimum characterizes the orbit as a whole. The fourth power enters only through the expansion Eq.~\eqref{eq:fourth}, and replacing it by the
$2\alpha$-th power carries every step for integer $\alpha\ge2$, with $\alpha$ factors $c_a(x_i)$ and $\alpha$
conjugates $\overline{c_a(y_j)}$ under the constraint $\bigoplus_i x_i\oplus\bigoplus_j y_j=0$, so $\ket{\sqrt p}$ minimizes $M_\alpha$ at every integer order.

\section{Numerical methods}\label{app:num}
\subsection{Optimization and statistics}
We evolve the states as full statevectors in double precision. The rotation angles are initialized
uniformly in $[0,2\pi)$ and optimized with Adam for $1000$ epochs, under a cosine-annealed learning rate
from $0.02$ to a tenth of that value, with global-norm gradient clipping at unit norm and reverse-mode
automatic differentiation. Probabilities entering the logarithm are regularized at $10^{-12}$.
A run is converged when its loss history has flattened, which happens well before the epoch budget ends
(Fig.~\ref{fig:twodfam} for the two-dimensional curriculum), and the converged loss quoted throughout is
the final-epoch value. The cost of exact SRE evaluation and of training confines the training study to
$\NQ\le12$; the entangling check of Sec.~\ref{sec:model} involves no training and reaches $\NQ=20$.

Every configuration is run $20$ times. Each family uses the same $20$ sets of initial angles, with an
independent entangler drawn for each run of the random families, so that a run of one family is paired with
a run of every other. For the spin glass we use $10$ disorder realizations with $10$ runs each at $L=12$, the same $10$ entangler draws serving every realization and temperature,
and five runs each at the auxiliary depths. Per-family losses are the mean and standard deviation over
runs, the median over the runs of a disorder realization being taken first for the spin glass.
Figure~\ref{fig:m2scan} reports the median and the interquartile range, and Fig.~\ref{fig:twoddist} the
lowest-loss run. Paired differences between families are formed on the shared initial angles and target,
and on the shared disorder realization for the spin glass, and are the mean and standard error over those.
Entanglement is the von Neumann entropy of the left $\lfloor\NQ/2\rfloor$ qubits. For the random families
the probe-state SRE is averaged over the same $20$ draws used for training, and the kicked-Ising family is
a single fixed gate.

For the random-phase references of Fig.~\ref{fig:gibbskld}(b) we take, for each temperature and each of the
$10$ disorder realizations, the lowest-loss Clifford-entangler run, assign independent phases uniform in
$[0,2\pi)$ to $\sqrt{q_{\bm\theta}(x)}$ and to $\sqrt{p_\beta(x)}$, and evaluate the SRE of the resulting
states, averaging over three phase draws.

\subsection{Targets at prescribed nonstabilizerness}
The targets of Sec.~\ref{sec:m2scan} are obtained by minimizing $(\Mtwo(\ket{\sqrt p})-m^{*})^{2}$ over
$\bm\vartheta$ with $p=\mathrm{softmax}(\bm\vartheta)$, using Adam at learning rate $0.05$ for $2500$ steps
from two or three random starts and keeping the one closest to $m^{*}$. Requested floors
$m^{*}=0,\dots,7$ are met at the precision quoted, and $m^{*}=8$ converges to $7.40$.

\subsection{Training the two-dimensional targets}
For the two-dimensional targets the circuit is trained at increasing depth $L=1,2,4,6,8,10,12$. At each new
depth, $17$ of the $20$ runs are warm-started, the first $L_{\mathrm{prev}}$ layers taking the angles of the
lowest-loss run at the previous depth and the added layers starting from Gaussian angles of small standard
deviation. The remaining three runs start from uniformly random angles.

\bibliography{refs}
\end{document}